\documentclass[aps,prl,preprint,superscriptaddress,amsmath,amssymb]{revtex4}

\newcommand {\CA}{Cd$_{3}$As$_{2}$}

\newcommand {\CZA}{(Cd$_{1-x}$Zn$_{x}$)$_{3}$As$_{2}$}
\newcommand {\RHall}{$R_{13,24}$}
\newcommand {\Rtwo}{$R_{26,26}$}
\newcommand {\Rfour}{$R_{58,67}$}

\usepackage{graphicx}% Include figure files
\usepackage{dcolumn}% Align table columns on decimal point
\usepackage{bm}% bold math
\usepackage{color}
\usepackage{ulem}
\usepackage{txfonts}

\begin{document}

%\preprint{}


\title{Edge and bulk states in Weyl-orbit quantum Hall effect\\as studied by Corbino measurements}



\author{Yusuke Nakazawa}
\affiliation{Department of Applied Physics and Quantum-Phase Electronics Center (QPEC), the University of Tokyo, Tokyo 113-8656, Japan}

\author{Ryosuke Kurihara}
\affiliation{Institute for Solid State Physics (ISSP), the University of Tokyo, Kashiwa 277-8581, Japan}
\affiliation{RIKEN Center for Emergent Matter Science (CEMS), Wako 351-0198, Japan}

\author{Masatoshi Miyazawa}
\affiliation{Department of Applied Physics and Quantum-Phase Electronics Center (QPEC), the University of Tokyo, Tokyo 113-8656, Japan}

\author{Shinichi Nishihaya}
\affiliation{Department of Applied Physics and Quantum-Phase Electronics Center (QPEC), the University of Tokyo, Tokyo 113-8656, Japan}
\affiliation{Department of Physics, Tokyo Institute of Technology, Tokyo 152-8550, Japan}

\author{Markus Kriener}
\affiliation{RIKEN Center for Emergent Matter Science (CEMS), Wako 351-0198, Japan}

\author{Masashi Tokunaga}
\affiliation{Institute for Solid State Physics (ISSP), the University of Tokyo, Kashiwa 277-8581, Japan}
\affiliation{RIKEN Center for Emergent Matter Science (CEMS), Wako 351-0198, Japan}

\author{Masashi Kawasaki}
\affiliation{Department of Applied Physics and Quantum-Phase Electronics Center (QPEC), the University of Tokyo, Tokyo 113-8656, Japan}
\affiliation{RIKEN Center for Emergent Matter Science (CEMS), Wako 351-0198, Japan}

\author{Masaki Uchida}
\email[Author to whom correspondence should be addressed: ]{m.uchida@phys.titech.ac.jp}
\affiliation{Department of Applied Physics and Quantum-Phase Electronics Center (QPEC), the University of Tokyo, Tokyo 113-8656, Japan}
\affiliation{Department of Physics, Tokyo Institute of Technology, Tokyo 152-8550, Japan}


\date{\today}

\begin{abstract}
We investigate edge and bulk states in Weyl-orbit based quantum Hall effect by measuring a Corbino-type device fabricated from a topological Dirac semimetal {\CZA} film.
Clear quantum Hall plateaus are observed when measuring one-sided terminals of the Corbino-type device.
This indicates that edge states of the Weyl-orbit quantum Hall effect form closed trajectories consisting of Fermi arcs and chiral zero modes independently on inner and outer sides.
On the other hand, the bulk resistance does not diverge at fields where the quantum Hall plateau appears, suggesting that the Weyl orbits in the bulk region are not completely localized when applying electric current through the bulk region.
\end{abstract}

%\keywords{}

\maketitle



% If in two-column mode, this environment will change to single-column
% format so that long equations can be displayed. Use
%\begin{widetext}
%\end{widetext}

%\section{Introduction}
Topological Dirac and Weyl semimetals host unique surface states termed Fermi arcs, which discontinuously start and end at Weyl nodes in the momentum space. Under magnetic field, surface carriers possibly propagate between the Fermi arcs of opposite surface via a chiral zero mode which is a bulk Landau level with one-dimensional dispersion parallel to the field direction, and then complete a closed magnetic orbit called Weyl orbit \cite{NatCommun2014Potter, SciRep2016Zhang, NatRevPhys2021Zhang, RMP2021Lv}. {\CA} is a typical three-dimensional topological Dirac semimetal and systematic studies have led to a greater understanding of its quantum transport \cite{PRB2013Wang, NatMat2014Liu, PRL2014Borisenko, NatRevPhys2021Zhang}. Observations of the surface quantum Hall effect in {\CA} thick films ($t = 70 - 100$ nm) \cite{NatCom2019Nishihaya, NatCom2021Nishihaya} and nanoplates ($t = 60 - 80$ nm) \cite{NatCommun2017Zhang, Nature2019Zhang, PRL2019Lin} have supported the presence of the Weyl orbit and stimulated further studies on a new type of dissipationless electric current and nonlocal quantum correlations \cite{PRRes2020Zhe}. Recent theoretical studies have predicted the conductance quantization in the Weyl-orbit QHE by calculating the energy spectrum of the bulk chiral Landau levels and their Chern numbers \cite{PRB2021Nguyen, PRB2021Chang}. Detailed pictures of the edge mode in the Weyl-orbit QHE have been also theoretically proposed \cite{PRL2020Li, PRL2021Chen, PRL2017Wang}. Experimentally, as a distinct feature of the Weyl-orbit QHE, intrinsic coupling between top and bottom surface states has been demonstrated by dual-gating experiments of {\CA} thick films \cite{NatCom2021Nishihaya}. This is in stark contrast to topological insulator surface states, where cyclotron orbits are independently formed on top and bottom surfaces \cite{NatPhys2014Xu, NatCom2015Yoshimi}. Nevertheless, the details of how the carriers on the Weyl orbit travel through the edge and bulk states are still elusive. In this context, Corbino measurements are expected to be suitable for investigating such edge and bulk states in the Weyl-orbit QHE \cite{PRB1992Dolgopolov, PRB1995Jeanneret}.

In this study, we performed high-field transport measurements of the Corbino-type device fabricated from a topological Dirac semimetal {\CZA} film as shown in Fig. 1. Clear quantum Hall plateaus are observed in {\RHall}, indicating that the closed corner edge mode trajectories of the Weyl-orbit QHE are formed on the inner and outer sides. Here subscripts denote configurations of current and voltage terminals; for example, $R_{13,24} = V_{24}/I_{13}$ represents that the voltage is measured between 2 and 4 with flowing the current from 1 to 3. On the other hand, {\Rtwo} does not diverge at the quantization fields, suggesting that the Weyl orbits in the bulk region are not completely localized under the bias, which is in sharp contrast to the conventional 2D QHE.

\begin{figure}
\begin{center}
\includegraphics[width = 12 cm]{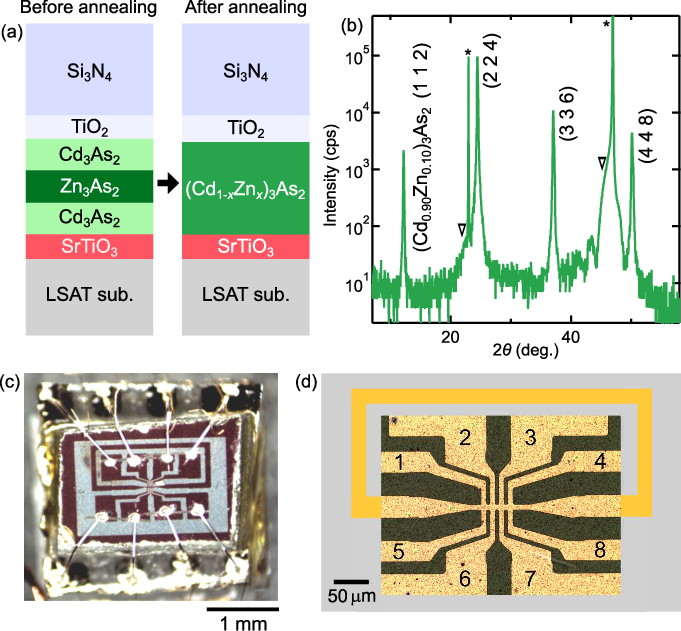}%
\caption{\label{fig:Fig1}
(a) Crosssectional view of the {\CZA} sample before and after annealing. (b) X-ray diffraction pattern of the sample after the annealing process. An asterisk and a triangle denote peaks from the LSAT substrate and the SrTiO$_3$ buffer layer, respectively. (c) Optical microscope image of the Corbino-type device mounted on a high-field transport measurement probe and (d) its magnified view around the center of the Corbino pattern with terminals 1 - 8.
}
\end{center}
\end{figure}

%\section{Experiment}
A Zn-doped {\CA} film with a thickness of 70 nm was grown by combining pulsed laser depositions and subsequent thermal annealing with capping layers TiO$_{2}$ and Si$_{3}$N$_{4}$ as detailed in previous papers \cite{NatCom2017Uchida, SciRep2018Nakazawa, SciAdv2018Nishihaya, PRB2018Nishihaya, PRB2019Uchida}. As shown in Fig. \ref{fig:Fig1}(a), one new approach here is to use (001) (LaAlO$_{3}$)$_{0.3}$(SrAl$_{0.5}$Ta$_{0.5}$O$_{3}$)$_{0.7}$ (LSAT) substrate and grow a few nm of SrTiO$_{3}$ buffer layer instead of using a (001) SrTiO$_{3}$ substrate as done in our previous works. This is because oxygen deficiency which affects transport measurements is inevitably introduced in the SrTiO$_{3}$ substrates during lithography processes. Zn-doping is for reducing the electron density of {\CA} and its composition is set to $x = 0.10$ in {\CZA}, which is small enough to maintain the non-trivial topological Dirac semimetal phase \cite{NatCom2019Nishihaya, PRB2018Nishihaya}. The carrier density is calculated to be $n_{\text{Hall}} = 1.7 \times 10^{12}$ cm$^{-2}$ from the low-field-slope of the Hall resistance. Figure \ref{fig:Fig1}(b) shows an x-ray diffraction $\theta - 2\theta$ scan of the film, confirming a (112)-oriented single phase. As shown in Figs. \ref{fig:Fig1}(c) and \ref{fig:Fig1}(d), the film was patterned into Corbino geometry by standard photolithography. High-field transport measurements were performed using a nondestructive pulsed magnet with a pulse duration of 37 ms at the International MegaGauss Science Laboratory at the Institute for Solid State Physics of the University of Tokyo. In the high-field measurements, DC excitation current of 10 $\mu$A was applied by Keithley 2400 sourcemeter. The voltage signals were measured by Yokogawa DL850E ScopeCorder after amplified by Stanford Research Systems SR560 preamplifier. Transport measurements up to 9 T were also performed using Quantum Design Physical Property Measurement System.

\begin{figure}
\begin{center}
\includegraphics[width = 11 cm]{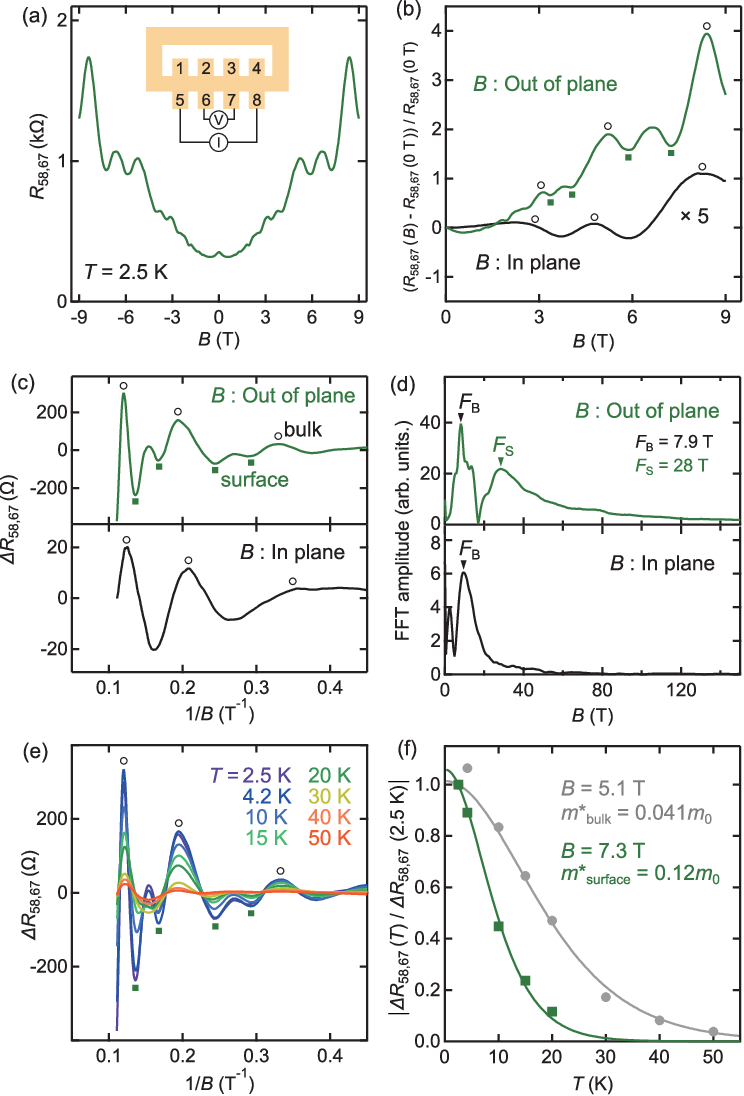}%
\caption{\label{fig:Fig2}
(a) Four-terminal magnetoresistance $R_{58,67} = V_{67} /I_{58}$ taken for out-of-plane magnetic fields at $T$ = 2.5 K. (b) Comparison of {\Rfour} between the out-of-plane (green) and in-plane (black) fields. Magnetic field is applied perpendicular to the current for the in-plane configuration. Filled squares denote quantum oscillations ascribed to the surface states, while open circles denote the 3D bulk quantum oscillations. (c) Oscillation components $\Delta${\Rfour} extracted by subtracting smooth backgrounds and (d) their Fourier transformations for the out-of-plane and in-plane data. Temperature dependence of (e) the out-of-plane oscillation components and (f) normalized oscillation amplitudes at typical fields. The surface and bulk effective masses are extracted from the Lifshitz-Kosevich formula.
}
\end{center}
\end{figure}

%\section{Results}
First, we study fundamental transport properties of the {\CZA} Corbino-type device by measuring four-terminal magnetoresistance {\Rfour} at low magnetic fields as shown in Figs. \ref{fig:Fig2}(a) and \ref{fig:Fig2}(b). Clear quantum oscillations are commonly observed for out-of-plane and in-plane fields, ensuring three dimensionality of the bulk Fermi surface in thick films \cite{NatCom2019Nishihaya, NatCom2021Nishihaya, APLMat2019Nakazawa}. For the out-of-plane field, additional oscillations marked by closed squares are observed, which indicate contributions from the surface states. Oscillation components $\Delta${\Rfour} are plotted for $1/B$ in Fig. \ref{fig:Fig2}(c), and their Fourier transformations are shown in Fig. \ref{fig:Fig2}(d). While the bulk oscillation component $F_{\text{B}}$ is clearly confirmed both for out-of-plane and in-plane fields, the surface one $F_{\text{S}}$ with higher frequency is observed only for the out-of-plane field configuration.

Temperature dependence of the oscillation amplitude $\Delta${\Rfour} is shown in Fig. \ref{fig:Fig2}(e) and theoretical fittings of the normalized amplitudes are plotted in Fig. \ref{fig:Fig2}(f). From the bulk oscillations, the effective mass is estimated at $m^{\text{*}}_{\text{bulk}} = 0.041m_{\text{0}}$ with the free electron mass $m_{\text{0}}$. The obtained value is in good agreement with previously reported ones for {\CA} bulk samples \cite{PRL2015Narayanan, PRX2015Zhao}. From the two-dimensional surface oscillations, in contrast, the effective mass is derived to be $m^{\text{*}}_{\text{surface}} = 0.12m_{\text{0}}$. This large effective mass compared to the bulk one is a hallmark of the intrinsic surface state \cite{NatCom2019Nishihaya, NatCommun2017Zhang}. The quantum scattering time of $\tau_{\text{q}} = 1.1 \times 10^{-13}$ s and the quantum mobility of $\mu_{\text{q}} = 3.1 \times 10^{3}$ cm$^{2}$/Vs are estimated from the Dingle analysis for the surface state \cite{Shoenberg1984}. From the above oscillation analysis, the surface quantum transport which evolves into the Weyl-orbit quantum Hall state is indeed confirmed in the Corbino-type device sample.

\begin{figure}
\begin{center}
\includegraphics[width = 12 cm]{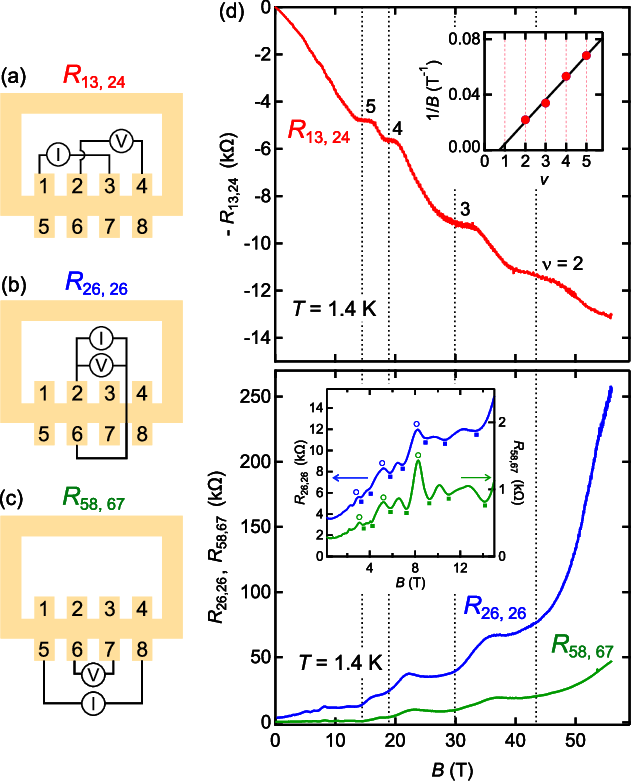}%
\caption{\label{fig:Fig3}
High-field measurements of the Corbino-type device for three different configurations (a) {\RHall}, (b) {\Rtwo}, and (c) {\Rfour}.(d) {\RHall} measured at $T$ = 1.4 K, showing clear quantum Hall plateaus supporting the edge picture. The inset shows filling factors plotted against the inverse of the magnetic field. (e) {\Rtwo} in the Corbino geometry measured at $T$ = 1.4 K. Four-terminal magnetoresistance {\Rfour} is also plotted for comparison. The inset magnifies the low field data. Filled squares and open circles denote the surface and bulk quantum oscillations.
}
\end{center}
\end{figure}

We next move on to high-field Corbino measurements of the Weyl-orbit quantum Hall state. The measurements were performed for three typical configurations {\RHall}, {\Rtwo}, and {\Rfour}, as illustrated in Figs. \ref{fig:Fig3}(a) - \ref{fig:Fig3}(c). As shown in Fig. \ref{fig:Fig3}(d), quantum Hall plateaus are observed in {\RHall}, verifying that the corner edge mode of the Weyl-orbit QHE forms closed trajectories on the inner edge of the Corbino-type device. The filling factors $\nu$ are linearly scaled by the inverse of the quantization fields as shown in the inset, which is consistent with the Weyl-orbit QHE observed in the standard Hall bar geometry \cite{NatCommun2017Zhang, NatCom2019Nishihaya}, and its slope gives the carrier density of $n_{\text{QHE}} = (e/h)\cdot(\Delta 1/B)^{-1} = 1.5 \times 10^{12}$ cm$^{-2}$, where $e$ is the elementally charge and $h$ is the Planck constant. The correspondence of $n_{\text{Hall}}$ and $n_{\text{QHE}}$ supports that both the bulk and surface states are involved in the formation of the Weyl-orbit quantum Hall state and the degeneracy of the Weyl orbit is fully lifted at the high magnetic fields \cite{NatCom2019Nishihaya, NatCom2021Nishihaya, NatCommun2017Zhang}.

Figure \ref{fig:Fig3}(e) shows {\Rtwo} and {\Rfour} with the same field scale as {\RHall}. Importantly, divergence behavior as discussed in detail later is not observed in {\Rtwo}. Instead, oscillatory behavior similar to the four-terminal resistance {\Rfour} is observed. These observations indicate that the bulk states of the Weyl-orbit QHE can be conductive under the bias, while the edge mode is well developed. This is in stark contrast to the conventional 2D QHE, where the bulk states are entirely localized in cyclotron motions. Here we need to point to non-zero $\sigma_{\text{xx}}$ by remaining bulk parallel conduction as a cause of non-divergence behavior in {\Rtwo}. However, such divergence behavior starts to appear for the conventional 2D quantum Hall system even when the bulk state is not completely localized and $R_\text{xx}$ minima does not reach zero resistance \cite{PRB1995Jeanneret}. This suggests that the observed non-divergence behavior of {\Rtwo} in our Corbino-type device may be related to a specific feature of the Weyl orbit. Given that the observed conductive bulk behavior originates from the carrier transfer between adjacent Weyl orbits, the transfer process is more likely to occur at around the surface Weyl nodes than in the bulk tunneling process because most of the density of states (DOS) of the Weyl orbits is located near the top and bottom surfaces and the DOS rapidly decays towards the bulk \cite{PRRes2020Zhe}.

\begin{figure}
\begin{center}
\includegraphics[width = 17 cm]{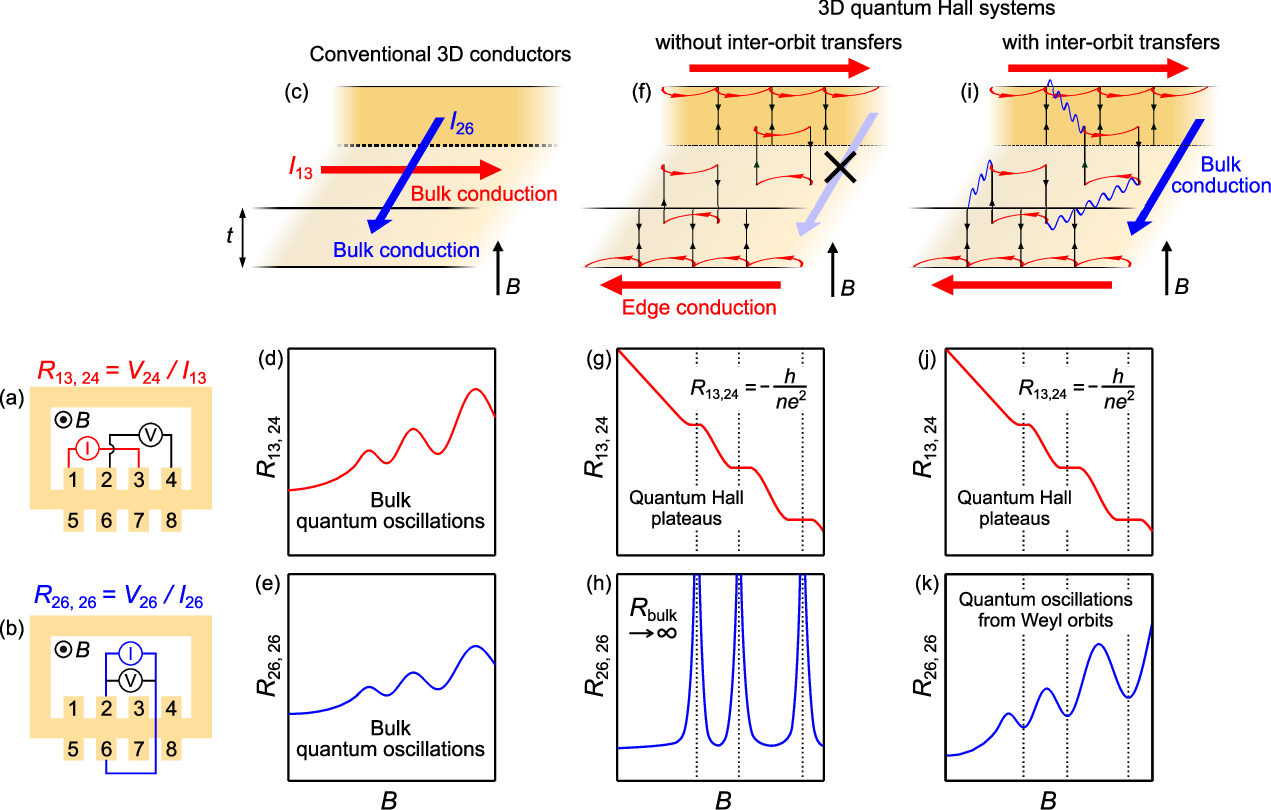}
\caption{\label{fig:Fig4}
Corbino measurements and expected magnetotransport for conventional 3D conductors and 3D quantum Hall systems based on Weyl orbits. Sketches of two configurations of current and voltage terminals are shown for measuring (a) {\RHall} and (b) {\Rtwo}. (c) - (e) The current is carried through bulk states in conventional 3D conductors, and Shubnikov-de Haas oscillations reflecting the bulk Fermi surface appear in both configurations. (f) - (h) In the absence of transfers between the Weyl orbits, current is carried only through the corner edge states consisting of Fermi arcs (red) and chiral zero modes (black), giving rise to a completely localized bulk state. In this case, quantum Hall plateaus are observed for {\RHall}, while {\Rtwo} diverges at quantization magnetic fields. (i) - (k) If carrier transfers can occur between the Weyl orbits, bulk conduction is allowed when applying current through the bulk region, and quantum oscillations are expected to appear for {\Rtwo}.
}
\end{center}
\end{figure}

%\section{Discussion}
Figures \ref{fig:Fig4} summarizes magnetotransport expected for conventional 3D conductors and 3D quantum Hall systems in Corbiono measurements. In conventional quantum Hall systems with edge channels, {\RHall} should be quantized as $R_{\text{yx}}$ measured in the Hall bar geometry. In a standard Hall bar, however, edge mode trajectories are not closed due to the presence of current electrodes. For the quantization of {\RHall}, on the other hand, it is essential that the edge modes are developed on the entire inner side to form a closed trajectory. Namely, by measuring {\RHall} in the Corbino-type device, it can be verified whether the edge states in the Weyl-orbit QHE form closed trajectories consisting of Fermi arcs and chiral zero modes. More precisely, current electrodes 1 and 3 are present even in this Corbino-type device. However, these electrodes may support only one-way propagation from top to bottom (or from bottom to top) surfaces in both 1 and 3 electrodes since they are attached on the same side of the film, and it can be concluded that ``another-way" propagation occurs through the chiral zero mode without electrodes on the opposite side of the inner edge. Another advantage of adopting the Corbino-type device is that bulk resistance in the Weyl-orbit QHE can be studied by measuring two-terminal resistance such as {\Rtwo}, since the inner and outer side edges are completely isolated in the Corbino-type device and contributions from the edge modes can be excluded.

Figures \ref{fig:Fig4}(c) - \ref{fig:Fig4}(k) schematize {\RHall} and {\Rtwo} expected for conventional 3D conductors and Weyl-orbit quantum Hall states. For conventional 3D conductors, Shubnikov-de Haas oscillations from the bulk Fermi surface are expected to be observed for both {\RHall} and {\Rtwo}, as shown in Figs. \ref{fig:Fig4}(c) - \ref{fig:Fig4}(e). In case the edge picture holds for the Weyl-orbit QHE, {\RHall} is expected to exhibit Hall plateaus as shown in Fig. \ref{fig:Fig4}(g), where closed trajectories of the edge mode develop as proposed in Ref. \cite{PRL2020Li}. According to this theoretical proposal \cite{PRL2020Li}, the edge mode in the Weyl-orbit QHE is completed by the Fermi arcs and the bulk chiral modes, as depicted in Fig. \ref{fig:Fig4}(f). Carriers traveling along the Fermi arc on the top surface transit into the bulk chiral mode at the Weyl node. After the carriers propagated through the bulk and reach the bottom surface, they are scattered into the bulk chiral mode again and bounced back to the top surface. This is because the same semicircle motion as on the top surface cannot be formed on the bottom surface due to the sample boundary. The carriers repeat these motions, resulting in the formation of conducting channels along the corner edge of the sample. Depending on sample orientation or arrangements of the Weyl nodes, Fermi arcs formed on side surfaces may also contribute to the formation of the corner edge conduction channels \cite{PRL2020Li}. Such complete motion of the corner edge mode has been first demonstrated in the present Corbino measurement of {\RHall}, while it was hindered due to the presence of current electrodes in previous conventional Hall bar experiments.

Regarding bulk resistance as measured by two-terminal configuration such as {\Rtwo}, two different cases can be considered depending on the nature of a set of propagation processes in the localized Weyl orbits. One case is that carriers are completely localized within a single Weyl orbit in the bulk region (Fig. \ref{fig:Fig4}(f)). In this case, electric current does not flow between the terminals 2 and 6 ($I_{\text{26}} = 0$) since the inner and outer side edges of the Corbino-type device are electrically isolated due to the complete localization of the Weyl orbits, while finite $V_{\text{26}}$ reflecting the potential difference between the inner and outer side edges is expected to be observed as in the conventional two-dimensional quantum Hall state \cite{PRB1995Jeanneret}. This results in divergence of {\Rtwo} at quantization magnetic fields (Figs. \ref{fig:Fig4}(g) and \ref{fig:Fig4}(h)), and the bulk region becomes highly insulating as in the conventional two-dimensional quantum Hall state. Another case is that carrier transfer between adjacent Weyl orbits is allowed especially when carriers start/end their semicircle cyclotron motion after/before propagating through the chiral zero mode (Fig. \ref{fig:Fig4}(i)). In this case, {\Rtwo} is expected to exhibit quantum oscillations reflecting formation of the Landau levels of the Weyl orbit \cite{NatCommun2014Potter} without divergence (Figs. \ref{fig:Fig4}(j) and \ref{fig:Fig4}(k)). One possible origin of the non-divergence behavior of {\Rtwo} in the present Corbino measurement is thus the inter-orbit carrier transfer, and further research is required for elucidating the incomplete localization of Weyl orbits.

%\section{Conclusion}
In summary, we have investigated the edge and bulk states of the Weyl-orbit QHE by measuring a Corbino-type device fabricated from the topological Dirac semimetal {\CZA}. Presence of closed trajectory of the edge states has been confirmed by the observation of quantum Hall plateaus for the one-sided terminals of the Corbino-type device. The bulk resistance measurement suggests that the bulk states of the Weyl-orbit QHE are not completely localized in stark contrast to the conventional 2D QHE. Our findings will stimulate further investigations for fully understanding the details of the carrier motion in Weyl-orbit QHE.

%\section{Acknowledgement}
This work was supported by JST PRESTO Grant Number JPMJPR18L2 and FOREST Program Grant Number JPMJFR202N, and by JSPS KAKENHI Grant Numbers JP22K18967, JP22H04958, JP22K20353, JP22H04471, JP21H01804, JP22H04501 from MEXT, Japan.

\newpage

\newpage

\newpage

\newpage


\begin{thebibliography}{100}

\bibitem{NatCommun2014Potter} A. C. Potter, I. Kimchi, and A. Vishwanath, Quantum oscillations from surface Fermi arcs in Weyl and Dirac semimetals, Nat. Commun. {\bf 5,} 5161 (2014).

\bibitem{SciRep2016Zhang} Y. Zhang, D. Bulmash, P. Hosur, A. C. Potter, and A. Vishwanath, Quantum oscillations from generic surface Fermi arcs and bulk chiral modes in Weyl semimetals, Sci. Rep. {\bf 6,} 23741 (2016).

\bibitem{NatRevPhys2021Zhang} C. Zhang, Y. Zhang, H.-Z. Lu, X. C. Xie, and F. Xiu, Cycling Fermi arc electrons with Weyl orbits, Nat. Rev. Phys. (2021).

\bibitem{RMP2021Lv} B. Q. Lv, T. Qian, and H. Ding, Experimental perspective on three-dimensional topological semimetals, Rev. Mod. Phys. {\bf 93,} 025002 (2021).

\bibitem{PRB2013Wang} Z. Wang, H. Weng, Q. Wu, X. Dai, and Z. Fang, Three-dimensional Dirac semimetal and quantum transport in Cd$_{3}$As$_{2}$, Phys. Rev. B {\bf 88,} 125427 (2013).

\bibitem{NatMat2014Liu} Z. K. Liu, J. Jiang, B. Zhou, Z. J. Wang, Y. Zhang, H. M. Weng, D. Prabhakaran, S.-K. Mo, H. Peng, P. Dudin, T. Kim, M. Hoesch, Z. Fang, X. Dai, Z. X. Shen, D. L. Feng, Z. Hussain, and Y. L. Chen, A stable three-dimensional topological Dirac semimetal Cd$_{3}$As$_{2}$, Nat. Mater. {\bf 13,} 677 (2014).

\bibitem{PRL2014Borisenko} S. Borisenko, Q. Gibson, D. Evtushinsky, V. Zabolotnyy, B. B{\"u}chner, and Robert J. Cava, Experimental Realization of a Three-Dimensional Dirac Semimetal, Phys. Rev. Lett. {\bf 113,} 027603 (2014).

\bibitem{NatCom2019Nishihaya} S. Nishihaya, M. Uchida, Y. Nakazawa, R. Kurihara, K. Akiba, M. Kriener, A. Miyake, Y. Taguchi, M. Tokunaga, and M. Kawasaki, Quantized surface transport in topological Dirac semimetal films, Nat. Commun. {\bf 10,} 2564 (2019).

\bibitem{NatCom2021Nishihaya} S. Nishihaya, M. Uchida, Y. Nakazawa, Y. Taguchi, and M. Kawasaki, Intrinsic coupling between spatially-separated surface Fermi-arcs in Weyl orbit quantum Hall states, Nat. Commun. {\bf 12,} 2572 (2021).

\bibitem{NatCommun2017Zhang} C. Zhang, A. Narayan, S. Lu, J. Zhang, H. Zhang, Z. Ni, X. Yuan, Y. Liu, J.-H. Park, E. Zhang, W. Wang, S. Liu, L. Cheng, L. Pi, Z. Sheng, S. Sanvito, and F. Xiu, Evolution of Weyl orbit and quantum Hall effect in Dirac semimetal {\CA}, Nat. Commun. {\bf 8,} 1272 (2017).

\bibitem{Nature2019Zhang} C. Zhang, Y. Zhang, X. Yuan, S. Lu, J. Zhang, A. Narayan, Y. Liu, H. Zhang, Z. Ni, R. Liu, E. S. Choi, A. Suslov, S. Sanvito, L. Pi, H.-Z. Lu, A. C. Potter, and F. Xiu, Quantum Hall effect based on Weyl orbits in {\CA}, Nature {\bf 565,} 331 (2019).

\bibitem{PRL2019Lin} B.-C. Lin, S. Wang, S. Wiedmann, J.-M. Lu, W.-Z. Zheng, D. Yu, and Z.-M. Liao, Observation of an Odd-Integer Quantum Hall Effect from Topological Surface States in {\CA}, Phys. Rev. Lett. {\bf 122,} 036602 (2019).

\bibitem{PRRes2020Zhe} Z. Hou and Q.-F. Sun, Nonlocal correlation mediated by Weyl orbits, Phys. Rev. Res. {\bf 2,} 023236 (2020).

\bibitem{PRB2021Nguyen} D.-H.-M. Nguyen, K. Kobayashi, J.-E. R. Wichmann, and K. Nomura, Quantum Hall effect induced by chiral Landau levels in topological semimetal films, Phys. Rev. B {\bf 104,} 045302 (2021).

\bibitem{PRB2021Chang} M. Chang and L. Sheng, Three-dimensional quantum Hall effect in the excitonic phase of a Weyl semimetal, Phys. Rev. B {\bf 103,} 245409 (2021).

\bibitem{PRL2020Li} H. Li, H. Liu, H. Jiang, and X. C. Xie, 3D Quantum Hall Effect and a Global Picture of Edge States in Weyl Semimetals, Phys. Rev. Lett. {\bf 125,} 036602 (2020).

\bibitem{PRL2021Chen} R. Chen, T. Liu, C. M. Wang, H.-Z. Lu, and X. C. Xie, Field-Tunable One-Sided Higher-Order Topological Hinge States in Dirac Semimetals, Phys. Rev. Lett. {\bf 127,} 066801 (2021).

\bibitem{PRL2017Wang} C. M. Wang, H.-P. Sun, H.-Z. Lu, and X. C. Xie, 3D Quantum Hall Effect of Fermi Arcs in Topological Semimetals, Phys. Rev. Lett. {\bf 119,} 136806 (2017).

\bibitem{NatPhys2014Xu} Y. Xu, I. Miotkowski, C. Liu, J. Tian, H. Nam, N. Alidoust, J. Hu, C.-K. Shih, M. Z. Hasan, and Y. P. Chen, Observation of topological surface state quantum Hall effect in an intrinsic three-dimensional topological insulator, Nat. Phys. {\bf 10,} 956 (2014).

\bibitem{NatCom2015Yoshimi} R. Yoshimi, A. Tsukazaki, Y. Kozuka, J. Falson, K. S. Takahashi, J.G. Checkelsky, N. Nagaosa, M. Kawasaki, and Y. Tokura, Quantum Hall effect on top and bottom surface states of topological insulator (Bi$_{1-x}$Sb$_{x}$)$_{2}$Te$_{3}$ films, Nat. Commun. {\bf 6,} 6627 (2015).

\bibitem{PRB1992Dolgopolov} V. T. Dolgopolov, A. A. Shashkin, N. B. Zhitenev, S. I. Dorozhkin, and K. von Klitzing, Quantum Hall effect in the absence of edge currents, Phys. Rev. B {\bf 46,} 12560 (1992).

\bibitem{PRB1995Jeanneret} B. Jeanneret, B. D. Hall, H.-J. B\"{u}hlmann, R. Houdr\'{e}, M. Ilegems, B. Jeckelmann, and U. Feller, Observation of the integer quantum Hall effect by magnetic coupling to a Corbino ring, Phys. Rev. B {\bf 51,} 9752 (1995).

\bibitem{NatCom2017Uchida} M. Uchida, Y. Nakazawa, S. Nishihaya, K. Akiba, M. Kriener, Y. Kozuka, A. Miyake, Y. Taguchi, M. Tokunaga, N. Nagaosa, Y. Tokura, and M. Kawasaki, Quantum Hall states observed in thin films of Dirac semimetal Cd$_{3}$As$_{2}$, Nat. Commun. {\bf 8,} 2274 (2017).

\bibitem{SciRep2018Nakazawa} Y. Nakazawa, M. Uchida, S. Nishihaya, M. Kriener, Y. Kozuka, Y. Taguchi, and M. Kawasaki, Structural characterisation of high-mobility Cd$_{3}$As$_{2}$ films crystallised on SrTiO$_{3}$, Sci. Rep. {\bf 8,} 2244 (2018).

\bibitem{PRB2018Nishihaya} S. Nishihaya, M. Uchida, Y. Nakazawa, K. Akiba, M. Kriener, Y. Kozuka, A. Miyake, Y. Taguchi, M. Tokunaga, and M. Kawasaki, Negative magnetoresistance suppressed through a topological phase transition in (Cd$_{1-x}$Zn$_{x}$)$_{3}$As$_{2}$ thin films, Phys. Rev. B {\bf 97,} 245103 (2018).

\bibitem{SciAdv2018Nishihaya} S. Nishihaya, M. Uchida, Y. Nakazawa, M. Kriener, Y. Kozuka, Y. Taguchi, and M. Kawasaki, Gate-tuned quantum Hall states in Dirac semimetal (Cd$_{1-x}$Zn$_{x}$)$_{3}$As$_{2}$, Sci. Adv. {\bf 4,} eaar5668 (2018).

\bibitem{PRB2019Uchida} M. Uchida, T. Koretsune, S. Sato, M. Kriener, Y. Nakazawa, S. Nishihaya, Y. Taguchi, R. Arita, and M. Kawasaki, Ferromagnetic state above room temperature in a proximitized topological Dirac semimetal, Phys. Rev. B {\bf 100,} 245148 (2019).

\bibitem{APLMat2019Nakazawa} Y. Nakazawa, M. Uchida, S. Nishihaya, S. Sato, A. Nakao, J. Matsuno, and M. Kawasaki, Molecular beam epitaxy of three-dimensionally thick Dirac semimetal Cd$_{3}$As$_{2}$ films, APL Mater. {\bf 7,} 071109 (2019).

\bibitem{PRL2015Narayanan} A. Narayanan, M. D. Watson, S. F. Blake, N. Bruyant, L. Drigo, Y. L. Chen, D. Prabhakaran, B. Yan, C. Felser, T. Kong, P. C. Canfield, and A. I. Coldea, Linear magnetoresistance caused by mobility fluctuations in $n$-doped {\CA}, Phys. Rev. Lett. {\bf 114,} 117201 (2015).
  
\bibitem{PRX2015Zhao} Y. Zhao, H. Liu, C. Zhang, H. Wang, J. Wang, Z. Lin, Y. Xing, H. Lu, J. Liu, Y. Wang, S. M. Brombosz, Z. Xiao, S. Jia, X. C. Xie, and J. Wang, Anisotropic Fermi Surface and Quantum Limit Transport in High Mobility Three-Dimensional Dirac Semimetal {\CA}, Phys. Rev. X {\bf 5,} 031037 (2015).

\bibitem{Shoenberg1984} D. Shoenberg, Magnetic Oscillations in Metals. (Cambridge University Press, Cambridge, 1984).

\end{thebibliography}
\end{document}